\documentclass{aa}  

\usepackage{graphicx}
\usepackage{txfonts}
\newcommand{\kms}{\,km\,s$^{-1}$}
\newcommand{\pab}{Pa$_{\rm\beta}$1.28}
\newcommand{\hb}{H$_{\rm\beta}$}
\newcommand{\hbo}{H${\rm\beta o}$}
\newcommand{\hg}{H$_{\rm\gamma}$}
\newcommand{\hgsigma}{H$_{\rm\gamma_{\sigma}}$}
\newcommand{\hd}{H$_{\rm\delta}$}
\newcommand{\ms}{M$_{\sun}$}

\newcommand{\aFe}{\mbox{$\mbox{[$\alpha$/Fe]}$}}

\defcitealias{eftekhari2021}{E21}
\defcitealias{eftekhari2022}{E22}
\defcitealias{beasley2008}{B08}
\usepackage{xcolor}
\usepackage{amssymb}
\usepackage[normalem]{ulem}
\begin{document}

   \title{Extragalactic globular cluster near-infrared spectroscopy:}

   \subtitle{II. The stellar population synthesis model zero-point problem extends to the near-infrared}

   \author{E. Eftekhari\inst{1,2,3} \and
           A. Vazdekis\inst{1,2} \and
           R. Riffel\inst{1,2,4} \and
           L. G. Dahmer-Hahn \inst{5} \and
           A. L. Chies-Santos\inst{4} \and
           M. A. Beasley\inst{1,2,6} \and
           A. Villaume\inst{7,8} \and
           E. Zanatta\inst{9}}

   \institute{Instituto de Astrof\'isica de Canarias, E-38200 La Laguna, Tenerife,               Spain \and
              Departamento de Astrof\'isica, Universidad de La Laguna, E-38205 La Laguna, Tenerife, Spain \and
              School of Astronomy, Institute for Research in Fundamental Sciences - IPM, 19395-5531, Tehran, Iran \and
              Instituto de Física, Universidade Federal do Rio Grande do Sul (UFRGS), Av. Bento Gonçalves, 9500, Porto Alegre, RS, Brazil \and
              Shanghai Astronomical Observatory, Chinese Academy of Sciences, 80 Nandan Road, Shanghai 200030, China \and
              Centre for Astrophysics and Supercomputing, Swinburne University, John Street, Hawthorn VIC 3122, Australia \and
              Waterloo Centre for Astrophysics, University of Waterloo, Waterloo, Ontario, N2L 3G1, Canada \and
              Department of Physics and Astronomy, University of Waterloo, Waterloo, Ontario N2L 3G1, Canada \and 
              Instituto de Astronomia, Geof\'isica e Ci\^encias Atmosf\'ericas, Universidade de S\~ao Paulo, 05508-900 S\~ao Paulo, Brazil}

   \date{Received September 15, 1996; accepted March 16, 1997}

 
  \abstract{Many recent studies have pointed out significant discrepancies between observations and models of stellar populations in the near-infrared (NIR). With current and future observing facilities being focused in this wavelength range, properly assessing and solving these issues is of utmost importance. Here, we present the first application of the extragalactic globular cluster (GC) near-infrared spectroscopy survey, and present evidence that these GCs reveal an age zero-point problem of stellar population synthesis (SPS) models. This problem has already been identified in the optical range for the GCs of the Milky Way. Such an issue arises when derived GC spectroscopic ages appear older than the Universe itself. We extend this discussion for the first time to the NIR, specifically using the Pa$_{\rm\beta}$ line at 1.28~microns. We focus on the GCs of the nearby Centaurus~A galaxy using their NIR spectra. This work broadens our understanding of the age zero-point problem and emphasises the necessity to revisit and refine SPS models, especially in the NIR domain.}


   \maketitle
%

\section{Introduction}

    The age zero-point problem presents a significant challenge in stellar population synthesis (SPS) models, notably in spectral indices analysis and the accurate age determination of old stellar populations such as globular clusters (GCs). This issue emerges as a discrepancy where the spectroscopically derived ages of these populations (> 10 Gyr; e.g. \citealt{strader2005, chies2011b}) appear older than the estimated age of the Universe. This problem was initially identified in the optical spectra of Milky Way (MW) GCs. For example, the study of the optical Balmer line \hg\ in the high-metallicity GC 47~Tucanae (47~Tuc) by \citet{gibson1999} highlighted this issue.

    Several hypotheses have been proposed to explain the age zero-point problem. One key area of discussion is the influence of horizontal branch (HB) stars and turn-off (TO) on Balmer lines. These lines are sensitive to the age of the stellar populations and are mainly contributed by the hottest stars, like those at the main sequence turn-off (MSTO) and the HB. In metal-rich regimes, old populations usually have an HB that is predominantly a red clump of cool stars ($T_{\rm eff} < 5000\,\rm K$) and therefore the hotter MSTO stars primarily contribute to the Balmer line. \citet{vazdekis2001} examined the impact of HB on the Balmer lines in 47~Tuc and found a minimal influence. However, in low-metallicity regimes ($< -1\,\rm dex$), the HB contains stars with temperatures comparable to the MSTO, leading to an age degeneracy, where very old, low-metallicity populations appear younger from their integrated light. In such metallicity regimes, blue HB stars can complicate the age zero-point problem by predicting stronger Balmer lines for oldest simple stellar populations (SSPs). At intermediate metallicities (-1.0 $\leq$ [Z/H] $\leq$ -0.5), where the relative contributions of the HB stars depend on the morphology of this evolutionary branch, \citet{mendel2007} found that while HB morphology does affect \hb, its impact on derived ages for most Galactic GCs (GGCs) remains relatively minor, regardless of the HB morphology used. Additionally, atomic diffusion (the movement of elements throughout the layers of a star) alters the surface composition and temperature of stars, especially at the TO point, making them appear cooler and thus older. Incorporating atomic diffusion into modelling isochrones can cause spectroscopic age estimates to move closer to those derived from colour--magnitude diagrams (CMDs) \citep{vazdekis2001}. Moreover, $\alpha$-enhanced isochrones affect the temperature and luminosity of stars in the TO point. According to \citet{vazdekis2001}, implementing $\alpha$-enhanced isochrones in SPS models leads to a closer alignment between spectroscopic age estimates and those derived from CMDs. 

    The nebular emission fill-in hypothesis also attempts to address the age zero-point problem; it proposes that gas emission might partially fill in Balmer lines, affecting their strengths and consequently the age estimates derived from them. However, the study of the behaviour of higher-order Balmer lines by \citet{vazdekis2001} reveals inconsistencies in the effects of emission fill-in across different Balmer lines, weakening the support for this hypothesis.

    Another aspect to consider in explaining the age zero-point problem is the impact of the systematic uncertainties influencing fundamental stellar atmospheric parameters, such as effective temperature, surface gravity, and iron abundance, on the line-strength predictions of SPS models.  The study by \citet{percival2009} suggests that typical systematic offsets of 100~K in T$_{\rm eff}$, 0.15~dex in [Fe/H], and 0.25~dex in log~g affect the \hb\ index's derived ages by up to 20\% for both old and intermediate-age populations. The interpretation of ages becomes more challenging when considering the H$_{\rm\delta F}$ and H$_{\rm\gamma F}$ indices, as their response to systematic shifts in [Fe/H] contrast with those of \hb. However, the study by \citet{vazdekis2010} indicates that overcoming the model zero-point problem requires a larger temperature offset of around 150~K, which exceeds current temperature scale uncertainties. These authors used line-strength diagnostic diagrams from their models and alternative models with different temperature scales to determine the ages and metallicities of GGC samples.

    All these findings point to the fact that interpreting Balmer lines is complicated, and different factors can influence GC age estimates. As another example, a recent study by \citet{leath2022} brings new insights into interpreting Balmer lines. Their work, which focuses on integrated spectroscopy of GGCs, points out a split in the \hbo\ line-strength index at intermediate to high metallicities ([Z/H] > -1), resulting in an apparent `upper branch' and  `lower branch' in the \hbo--[MgFe] diagram\footnote{These two branches do not show up at first sight when using the \hb\ index.}. Contrary to previous belief that this was caused by the presence of hot blue straggler stars, leading to an underestimation of spectroscopic ages \citep{cenarro2008}, their findings suggest that this split is actually due to increased helium abundance. 

    Interestingly, the age zero-point problem is also observed when using extragalactic GCs. For example, Fig.~8 in \citet{cenarro2007} shows that most metal-poor GCs in the elliptical galaxy NGC~1407 fall below model predictions for Balmer lines. Likewise, \citet{leath2022}'s exploration of He abundance effects on the \hbo\ index in M31 GCs shows a similar pattern (see their figure~11). Such studies expand our understanding of the age zero-point problem, offering insights into environmental impacts on stellar populations.

    An additional aspect of the age zero-point problem that is worth considering is the extension of the analysis to the near-infrared (NIR) spectral range. While past studies have mainly focused on optical wavelengths, recent advancements have shifted attention to the NIR spectral range. This range is often invoked to address optical challenges such as the age--metallicity degeneracy and high sensitivity to dust. However, SPS models in this spectral range are still in their infancy. This is particularly evident in the failure of SPS models to reproduce some NIR absorptions in galaxies and the challenges in interpreting CO absorption bands in massive early-type galaxies (ETGs) \citep{marmol2009, baldwin2018, dahmer2018, riffel2019, eftekhari2022}. Aiming to investigate these issues further, we observed the first data set of NIR spectra of extragalactic GCs (\citet{dahmer2024}, hereafter Paper~I). This sample is composed of 21 GCs in Centaurus~A/NGC~5128 (hereafter Cen~A). Cen~A is among the most massive galaxies in the nearby Universe, with an estimated stellar mass of 10$^{11.21}$~\ms\ \citep{fall2018}. It is classified as a lenticular galaxy \citep{devaucouler1991}, and its morphological features include a prominent dust lane, extensive lobes, and jets, which are indicative of its active galactic nucleus. The present study, part of the  extragalactic GC NIR spectroscopy survey, is the first to investigate the age zero-point problem in the NIR, displaying it in the distinct environment of Cen~A, thus providing new insights into NIR SPS modelling and the interpretation of age-sensitive indices.

    The paper is organised as follows: Section~\ref{sec:data} details the sample and data used in the study. Section~\ref{sec:models} describes the stellar population models employed. In Section~\ref{sec:analysis}, we describe the spectral analysis, outlining our methodologies for data homogenisation and index measurements. The results of our analysis are presented in Section~\ref{sec:results}. Finally, we provide a summary of our findings and conclusions in Section~\ref{sec:conclusions}. 

\section{Samples and data}\label{sec:data}

    Our investigation of the model zero-point problem makes use of different data sets, including NIR and optical spectra. In this section, we outline the selection criteria for GCs in the Cen~A galaxy, for which NIR spectra were obtained, and summarise pre-existing optical spectral data. Subsequently, we describe the NIR and optical datasets for MW GCs with which we compare the GCs in Cen~A. 

        \subsection{Centaurus~A GC data}

        We selected a sample of 21 GCs in the Cen~A galaxy based on the optical catalogue by \citet{beasley2008}, which was designed to cover as many ages and metallicities as possible. As an age proxy, we chose \hbo, and  we used the composite index [MgFe]$^{\prime}$ \citep{thomas2003} as our total metallicity indicator, which was constructed from the combination of three indices: Mg$_{b}$, Fe\,5270, and Fe\,5335. This index is insensitive to variations in the \aFe\ abundance ratio \citep{vazdekis2015}. The GCs were observed in the NIR using the TripleSpec4 ($\rm{ R=\lambda/\Delta\lambda\simeq3500}$) instrument on the SOAR telescope during observing runs in March and April 2020 and 2021. The NIR spectra were reduced following standard procedures. We employed the Spextool IDL pipeline of \citet{vacca2003, cushing2004} for data reduction, incorporating flat-field correction, wavelength calibration, spatial extraction, telluric band correction, and flux calibration. The final spectra were corrected for the Doppler shift of Cen~A and MW dust reddening, yielding rest frame spectra covering 0.96--2.46~microns. Further details of the observations and data reduction procedures are described in Paper~I.

        We also use the 21 optical spectra from the \citet{beasley2008} survey of Cen~A GCs. This survey was conducted using the 2-degree field instrument on the Anglo-Australian Telescope. This provided high-quality spectra over a wide field, which are essential for the metallicity, age, and kinematic analysis of the GC system. The spectra were obtained with a resolution of R$\approx$2000, covering the wavelength range of 3700--5800~\AA.

        \subsection{Milky Way GC data} 

        We incorporated the NIR spectroscopic dataset of MW GCs from \citet{riffel2011} for comparative purposes. These data were selected from the \citet{bica1986} catalogue, which was constructed in order to provide spectral observational constraints for calibrating evolutionary population models in the NIR. \citet{riffel2011} obtained cross-dispersed NIR spectra for 12 GCs using the Ohio State Infrared Imager/Spectrometer mounted on the SOAR Telescope, covering the 1.2~micron to 2.35~micron region with a resolving power of R$\sim$1200. Their observational strategy and instrumental configuration were optimised to minimise the impact of seeing and aperture variations, yielding a reliable subset of 12 GCs with S/N $\geq$ 25 and spanning $-1.62 \leq \rm [Z/H] \leq -0.04$ range. These clusters were originally chosen for their wide ranges of age and metallicity, which is relevant for testing evolutionary models. 

        We also performed a cross-match of 12 GGCs studied in \citet{riffel2011} with the optical spectroscopy datasets of \citet{schiavon2005}, \citet{kim2016}, and \citet{usher2017}, which yielded 11 matches. These datasets feature wide-ranging wavelength coverage and relatively high spectral resolution. The library of integrated spectra of MW and Local Group GCs by \citet{usher2017} covers a wavelength range of 3300~\AA\ to 9050~\AA\ with a resolution of R = 6800. The wavelength coverage of integrated spectra of 24 GCs in \citet{kim2016} is from 4000 to 5400~\AA\ with a spectral resolution of FWHM$\sim$2.0~\AA. The library of \citet{schiavon2005} covers the range $\sim$3350--6430~\AA\ with a resolution of $\sim$3.1~\AA\  and integrates the full projected area within cluster core radii. 

\section{Stellar population models} \label{sec:models}

    Stellar population models are required to interpret observations of stellar clusters and galaxies. As each model is constructed using its own assumptions, methods, and distinct ingredients, variations in their outputs can naturally be anticipated. To determine whether the age zero-point problem is prevalent across various sets of models, we employed different SPS models. In the following, we summarise the construction and characteristics of these models.

    \begin{itemize}

        \item Updated E--MILES \citep{rock2015, rock2016, vazdekis2016}: 
    
            The updated E--MILES models (Vazdekis et al. 2024, in prep.) are now used to make predictions  in the NIR that incorporate both the original \citep{cushing2004,rayner2009} and the Extended IRTF (E-IRTF) stellar library \citep{villaume2017}. These models mainly improve the predictions in the NIR for metallicities below -0.4~dex. E--MILES wavelength coverage extends from 1680 to 50000~\AA. These models employ theoretical isochrones from the BaSTI \citep{pietrinferni2004} and Padova \citep{girardi2000} databases converted to the observational plane by employing extensive photometric stellar libraries (see \citet{vazdekis2016} for a detailed description). The predictions in the NIR are  considered reliable for ages above 1~Gyr (although in the optical range, the predictions are considered reliable for much younger populations). The SSP spectral predictions have a resolution of FWHM = 2.5~\AA\ in the optical and R $\approx$ 2000 in the NIR. The E--MILES models also incorporate a variety of initial mass functions (IMFs), including the universal and revised Kroupa \citep{kroupa2001}, Chabrier \citep{chabrier2001}, and unimodal and bimodal IMFs \citep{vazdekis1996}. Here, we use the updated E--MILES model predictions based on the BaSTI(Padova) isochrones and consider ages ranging from 5 to 13~Gyr (5.01 to 12.59~Gyr) and total metallicities spanning from as low as -2.27 (-2.32) up to the solar value of +0.06 (0.00). We adopt a low-tapered ($<$0.6\ms) bimodal IMF with a logarithmic slope of 1.3, which almost resembles the standard Kroupa IMF (see \citet{vazdekis1996}).

        \item $\alpha$-enhanced E--MILES \citep{labarbera2017}

            In addition to the scaled-solar models, we also incorporate $\alpha$-enhanced E-MILES models, which are an updated version of the Na-enhanced models described in \citet{labarbera2017}, to examine the impact of varying $\alpha$-element abundances. For this study, we used two sets of models: one with [$\alpha$/Fe]=0 and another with [$\alpha$/Fe]=+0.4. Both sets of models were computed using the Padova isochrones. The first set ([$\alpha$/Fe] = 0) includes stellar populations with isochrones and a stellar library assuming solar-scaled abundances, covering ages from 5 to 13~Gyr and metallicities ranging from [Z/H]=-0.66 to +0.06. The second set ([$\alpha$/Fe] = +0.4) consists of $\alpha$-enhanced models with isochrones and a stellar library enhanced by [$\alpha$/Fe]=+0.4, covering the same age range but with metallicities from [Z/H]=-0.35 to +0.06.

    \end{itemize}

\section{Spectral analysis}\label{sec:analysis}

    In this section, we describe the key steps of our analysis. In Sect.~\ref{sec:homogenisation}, we show the procedures undertaken to ensure homogeneity and consistency across the different spectral datasets. The aim of this processing step is to prepare the data for subsequent line-strength measurements and analysis. In Sect.~\ref{sec:measurements}, we present the selected indices, their significance, and the methodology  adopted to measure them. 

    \subsection{Data-set homogenisation}\label{sec:homogenisation}

        \subsubsection{Centaurus~A GC spectra}
        
            \noindent The optical spectra of Cen~A GCs provided in the air wavelength system were in the observed frame. We therefore used the radial velocities of the GCs taken from \citet{beasley2008} to bring the original optical spectra to the rest frame. The spectral resolution of these observations varied, with FWHM=5.3~\AA\ for spectrograph A and 4.0~\AA\ for spectrograph B.

            The NIR spectra of Cen~A GCs were initially in the vacuum wavelength system, and therefore we transformed them to the air system by applying a conversion formula similar to the one used by SDSS. We then adjusted for the observed frame using the redshift of the Cen~A galaxy (z = 0.00183, \citep{graham1978}) and the radial velocities of GCs provided by \citet{beasley2008}, thereby placing these NIR spectra in the rest frame. In this study, our primary focus is on the \pab\ absorption feature. Upon closely comparing the redshifted spectra of the GCs with a reference SSP model, we found that the dip of the Pa$_{\beta}$ feature in most GCs does not align exactly with the model in terms of wavelength. To resolve this, we applied a second-order wavelength correction, shifting the spectra around the Pa$_{\beta}$ feature to ensure they were precisely in the rest frame. It is crucial to have the Pa$_{\beta}$ absorption accurately positioned at the rest frame for line-strength analysis, as \citet{eftekhari2021} demonstrated that this index is sensitive to wavelength shifts and an uncertainty of more than 92~\kms\ in wavelength calibration can lead to significant variation in the measured index value. Additionally, the NIR spectra of the Cen~A GCs include error spectra.

        \subsubsection{Milky Way GC spectra}

            The optical spectra of the MW GCs had already been converted to the air wavelength system. The spectra from \citet{schiavon2005} and \citet{kim2016} were also provided in the rest frame. However, the spectra from \citet{usher2017} were not. Therefore, we used the {\sc pPXF} code \citep{capellari2004,capellari2017} to derive the radial velocities of this sample and shift their spectra to the rest frame. The \citet{usher2017} and \citet{kim2016} samples include error spectra as well.

            The NIR spectra of MW GCs were originally provided in the air system and in the rest frame. However, similar to the NIR spectra of Cen~A GCs, the dip of the \pab\ absorption feature did not exactly match the model. Therefore, we performed a second-order wavelength calibration around this index by applying a spectral shift to align the feature more precisely.

        \subsubsection{Resolution homogenisation}
        
            \noindent Before proceeding with line-strength measurements, it was imperative to ensure that all spectra shared a common resolution. To achieve this, we employed a pixel-by-pixel convolution method (i.e convolving each pixel individually to achieve the desired resolution) originally derived from the {\sc pPXF} code \citep{capellari2004,capellari2017}. Specifically, all spectra in the optical data sets were convolved to a uniform resolution of $\sigma$=180~\kms, aligning them with the lowest resolution among the optical data. Similarly, for the NIR samples, we convolved all spectra to a resolution of $\sigma$=106~\kms, which corresponds to the lowest resolution among the NIR data. We note that for Cen~A GCs data sets, we considered the intrinsic broadening of spectral features for each GC by incorporating the velocity dispersion information provided by \citet{beasley2008}. This homogenisation process is essential, as broadening can affect line-strength measurements.

    Throughout these procedures, we carefully processed the spectra to ensure consistency across all data sets. As part of this process, we employed the {\sc SpectRes} Python tool \citep{carnall2017} to resample the spectra and their associated uncertainties onto an arbitrary wavelength grid.

    \subsection{Index measurements}\label{sec:measurements}

        In stellar population studies, hydrogen spectral lines play a crucial role in characterising the age of stellar populations. The Balmer line \hb, which corresponds to the transition from the n=2 to n=4 energy level in hydrogen at $\sim$4861~\AA, is highly sensitive to the age of stellar populations. Other Balmer lines in the optical window, such as \hg\ and \hd, are less used for spectroscopic age estimates as they show sensitivity to variations in the abundance of $\alpha$ elements. One notable exception is the \hgsigma\ index definition of \citet{vazdekis1999b} and \citet{vazdekis2001}, which requires extremely high signal-to-noise ratios \citep{vazdekis2015}, and these are not satisfied by our GC spectra. In our study, we employ the optimised index \hbo, which is derived in a manner that maximises its sensitivity to age while minimising its dependence on metallicity \citep{cervantes2009}. Therefore, the \hbo\ index is able to provide better age estimates for the GCs of Cen~A, enabling us to explore the age zero-point problem in extragalactic GCs.

\begin{figure}
        \centering
        \includegraphics[width=0.99\linewidth]{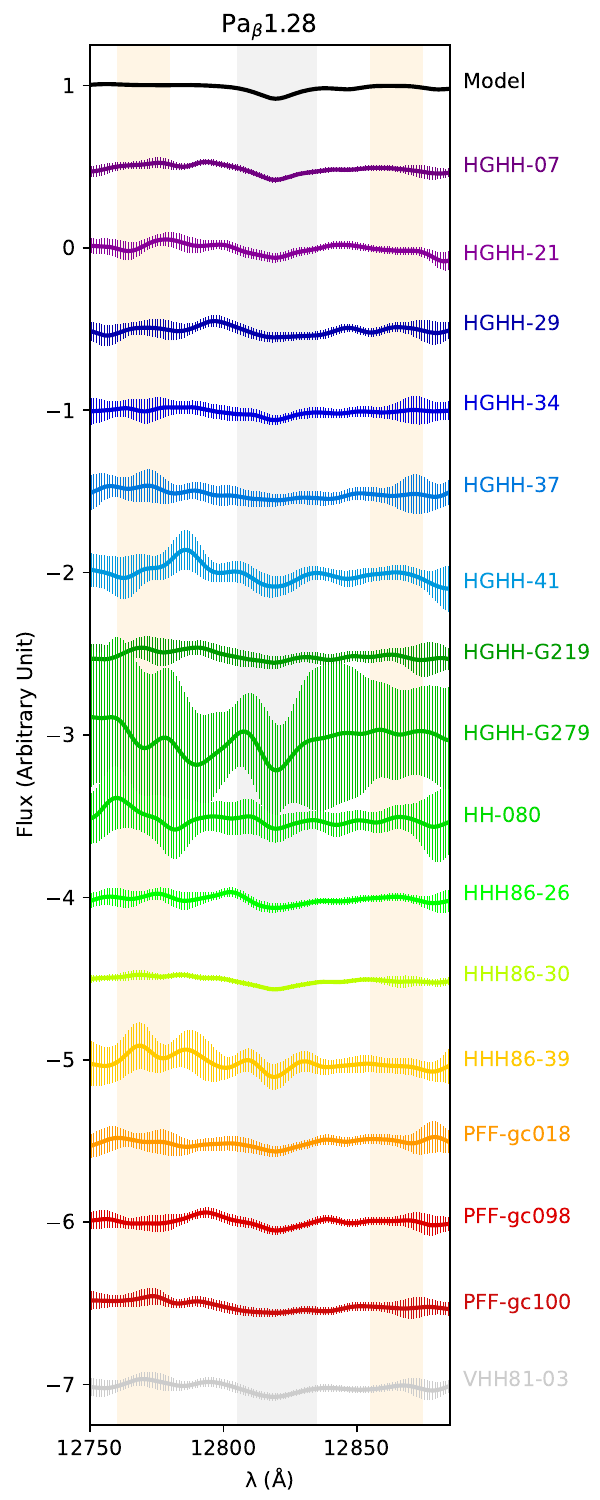}
        \protect\caption[]{Spectra of Cen~A GCs in the region of the \pab\ index. All spectra are presented in the rest frame and have been smoothed to a resolution of 106~\kms. Error bars are shown as vertical lines for all spectra. The black spectrum on top represents the E--MILES SSP of age 11~Gyr, at Z=-0.66, with a Kroupa-like IMF. The central bandpass of the \pab\ index, along with the blue and red pseudo-continua, are highlighted as grey and orange areas, respectively. These bandpasses are based on \citet{eftekhari2021}, with the central bandpass narrowed to 12805--12835~\AA\ to avoid spectral artifacts within the index.}
        \label{fig:fig1}
        \end{figure}

        To address the age zero-point problem in the NIR, we rely on the Paschen line centred at 1.28~microns. We use the index definition of \citet{eftekhari2021} as it is optimised to be maximally sensitive to the age of the stellar populations. Figure~3 in \citet{eftekhari2021} shows a remarkable drop in the \pab\ line strength with increasing age for populations older than 2~Gyr. Importantly, \pab\ is notably more responsive to changes in age than to variations in metallicity or the IMF. Furthermore, it should be noted that the \pab\ index exhibits a mild dependency on titanium abundance due to potential contamination of the central bandpass by \ion{Ti}{i} lines (see fig.~A1 in E21). Unfortunately, some of the spectra of our sample of Cen~A GCs exhibit some artifacts within the central bandpass of the \pab\ original index  definition. To minimise the impact of these artifacts, we narrowed the central bandpass to a range of 12805 to 12835~\AA. Despite this adjustment, five GCs still displayed artifacts within the central region of the \pab\ index and for that reason, we excluded them from our analysis. A zoom onto the \pab\ index in the spectra of the Cen~A GCs is shown in Fig.~\ref{fig:fig1}, where the pseudo-continua and central bandpasses are marked in orange and grey, respectively. The spectra in the plot are already de-redshifted to the rest frame and smoothed to a velocity dispersion of 106~\kms. It is worth noting that our study marks the first analysis of \pab\ absorption in the NIR for GCs, and a further novelty is that we are analysing GCs in an extragalactic context.

        We measured the line strengths of the aforementioned indices in both the observed and model spectra. Line-strength measurements were performed using the {\sc indexf} code, a software tool developed by \citet{cardiel2010}. As we have error spectra for Cen~A GCs in the NIR, as well as GGCs from \citet{usher2017} and \citet{kim2016} datasets, the {\sc indexf} code estimated line-strength errors using the provided error spectrum. The code employs error-propagation techniques ---accounting for the noise characteristics present in the error spectrum--- to determine the associated uncertainties in the line-strength measurements. For datasets where error spectra were unavailable, we used the {\sc indexf} program to estimate random errors in the line-strength indices through a simulation-based approach. The program generated synthetic error spectra based on the original data, assuming randomly generated S/Ns (default range: 1 to 100). Using these synthetic error spectra, {\sc indexf} performed 1000 Monte Carlo simulations (\texttt{simulsn=1000}), introducing random variations following a Gaussian distribution. The program assumes these errors represent 1$\sigma$ variations. For each simulation, the line-strength index was measured, the final index value was taken as the mean value of all simulations, and its standard deviation was taken as the uncertainty on the index line-strength value. It is important to note that for the GGCs common to the datasets of \citet{usher2017}, \citet{kim2016}, and \citet{schiavon2005}, we employed the average values of the indices. 

\section{Results}\label{sec:results}

\begin{figure}
    \centering
    \includegraphics[width=0.99\linewidth]{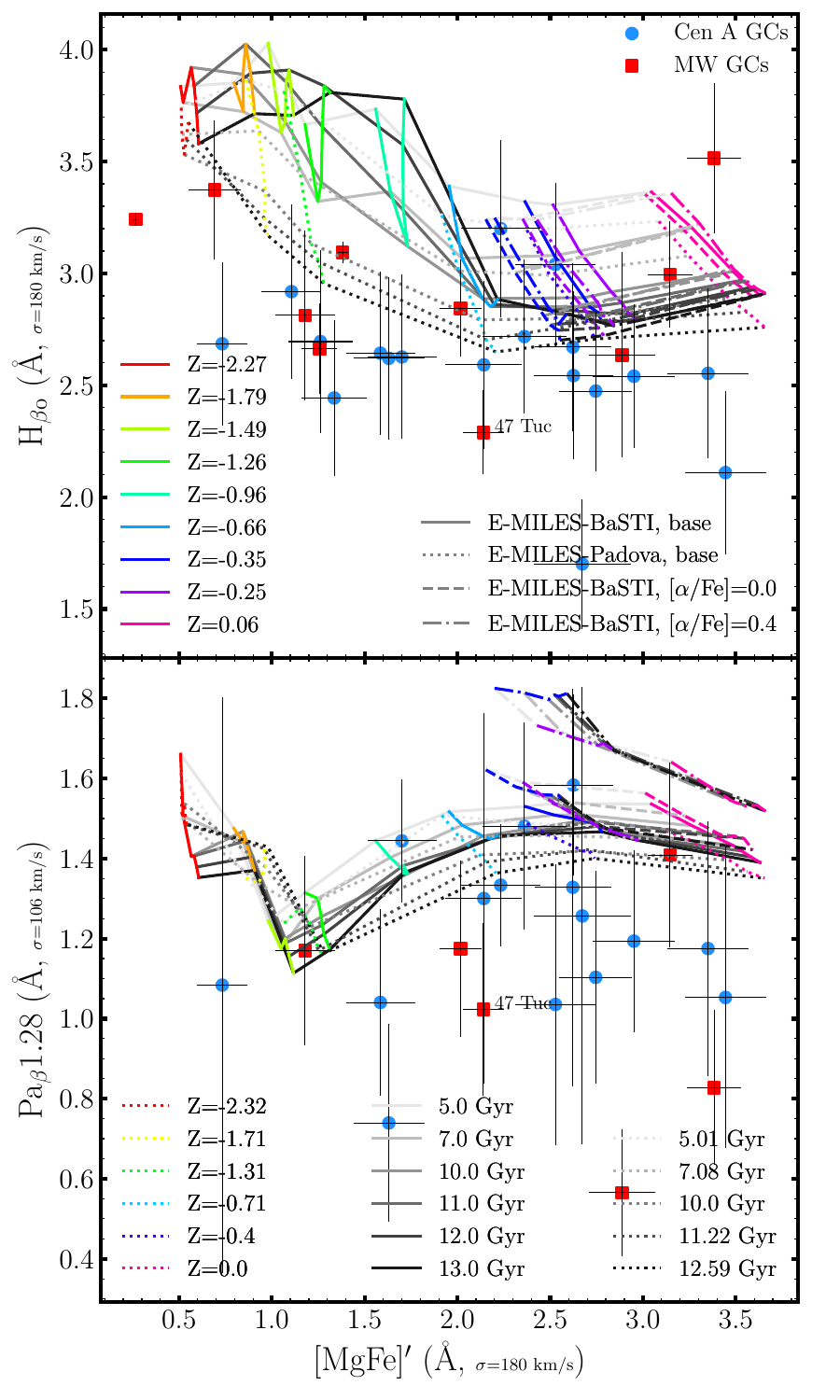}
    \protect\caption[]{Line strengths of two hydrogen line indices, \hbo\ in the optical (upper panel) and \pab\ in the NIR (lower panel), plotted against the total metallicity indicator [MgFe]$^{\prime}$ for two samples of GCs in the Cen~A galaxy (blue points) and MW (red squares). The overlaid grids represent predictions from updated E--MILES SSP models with BaSTI and Padova isochrones (solid and dotted lines), as well as E--MILES SSPs with [$\alpha$/Fe]= 0.0 and +0.4 (dashed and dash-dotted lines). These models span a range of metallicities from Z = -2.27 (-2.32) to solar, as indicated by various colours,  and ages ranging from 5 to 13~Gyr (12.59~Gyr) for SSPs with BaSTI (Padova) isochrones.}
    \label{fig:fig2}
    \end{figure}

    The model zero-point problem has been explored in the optical spectrum, primarily using MW GCs, but its origin remains a challenge to our understanding. Moving beyond the MW, we have expanded this research to include extragalactic GCs, focusing on those of the Cen~A galaxy. First, we measured the optical indices \hbo\ and [MgFe]$^{\prime}$, both expressed in \AA, for our sample of Cen~A GCs. The results are shown in the upper panel of Fig.~\ref{fig:fig2}, which shows the \hbo\ versus the total metallicity indicator. The blue points represent the measurements derived from the spectra of our sample of GCs in Cen~A. The solid grid represents updated E--MILES SSP models with BaSTI isochrones. These models have total metallicities from -2.27 to solar. Each colour on the grid represents a metallicity, as described in the legend. Model ages are shown with grey scale and span from 5 to 13~Gyr. It is evident that the GCs of Cen~A have a consistently lower \hbo\ value compared to the predictions of the E--MILES models with BaSTI isochrones (except for HGHH-34 and HHH86-30). The observed discrepancy between the measurements of the Cen~A GCs and our reference model confirms that the model zero-point problem also appears in extragalactic GC systems (e.g. \citet{brodie2005, cenarro2007, chies2012}).
    
    To extend our analysis of the model zero-point problem to the NIR, we focus on the \pab\ index. The lower panel in Fig.~\ref{fig:fig2} shows \pab\ line strengths against the total metallicity indicator, [MgFe]$^{\prime}$. As in the upper panel, the blue data points represent measurements from the spectra of Cen~A GCs. These points are overlaid with the predictions of the E--MILES SSP models with BaSTI isochrones for varying metallicities and ages (solid grid). Similar to the \hbo\ panel, a majority of the GCs in the Cen~A galaxy are situated below the model grid in the NIR. This observation emphasises that the age zero-point problem is not confined to the optical range and reveals limitations within SPS models when extended into the NIR domain as well. Finally, it is remarkable that in this spectral range the model zero-point problem emerges across the entire range of metallicities in our sample of GCs. This observation aligns with previous findings in high- and intermediate-metallicity GCs (e.g. in 47 Tuc GCs, \citet{gibson1999}, \citet{vazdekis2001}, and \citet{mendel2007}). \citet{vazdekis2001} also emphasised the importance of investigating this age gap in metal-poor GCs ($\rm [Z/H] \lesssim -0.9$) and the necessity of enhancing stellar spectral libraries to improve their predictive accuracy for lower metallicity ranges.

    It is also noteworthy that the age zero-point problem in both the optical and NIR is not exclusive to the E--MILES model. The XSL model, one of the latest SPS models with high resolution (R$\sim$10000) and broad wavelength coverage from NUV to NIR, also exhibits the age zero-point problem. A detailed comparison of predictions from three recent and widely used SPS models is presented in Appendix~\ref{sec:AppendixA}.

    To test the impact of varying the stellar evolutionary prescriptions used in the E-MILES model, we compared grids corresponding to two distinct sets of isochrones: BaSTI and Padova. The grid of E--MILES models based on BaSTI isochrones is shown with solid lines, while the grid based on Padova isochrones is represented by dotted lines in Fig.~\ref{fig:fig2}. For both indices, the Padova-based grid consistently lies lower than the BaSTI-based grid. The BaSTI-based model predicts a sharper increase in the \hbo\ line strength for metallicities of Z $<$ -0.66 compared to the Padova-based model at the oldest ages. However, the predictions for \pab\ show minimal differences between the two isochrones. Similar to the E--MILEs models with BASTI isochrones, the models with Padova isochrones show the age zero-point problem for \hbo\ and \pab\ indices. A detailed discussion of the impact of isochrones on the predictions of the XSL model for hydrogen lines compared to those of the E--MILES model is presented in Appendix~\ref{sec:AppendixB}.

    \begin{figure}
    \centering
    \includegraphics[width=0.99\linewidth]{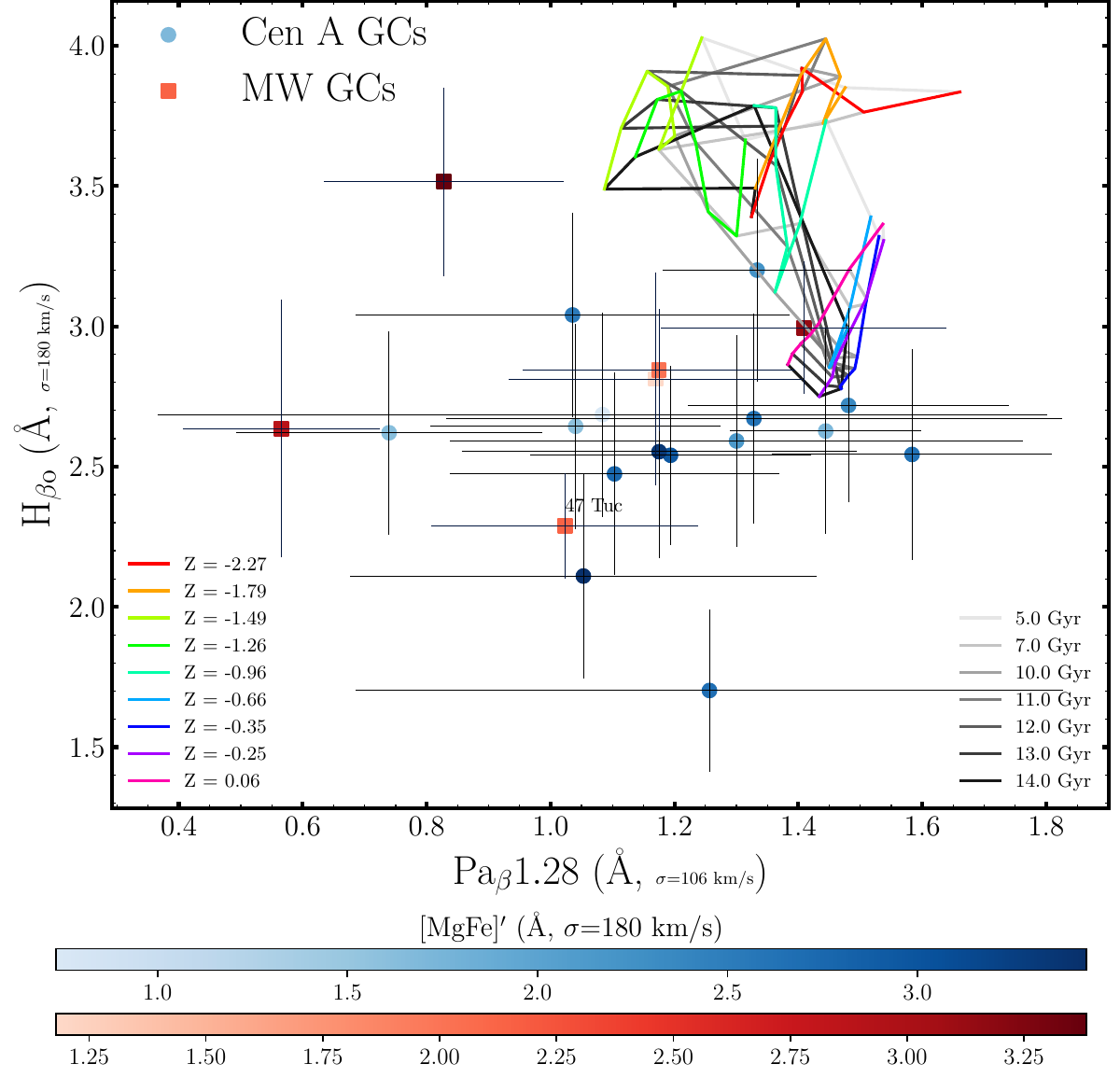}
    \protect\caption[]{\hbo\ vs \pab\ line strengths in GCs from the Cen~A galaxy and the MW. In this plot, Cen~A GCs are represented by filled circles and MW GCs by filled squares. The line strengths are plotted against the [MgFe]$^{\prime}$ metallicity indicator, with a bluish colour bar denoting the [MgFe]$^{\prime}$ values for Cen~A GCs and a reddish colour bar for MW GCs. The overlaid grid shows the predictions of the E--MILES model with BaSTI isochrones. The model grid displays predictions across a range of metallicities for the distinct ages specified in the legend.}
    \label{fig:fig3}
    \end{figure}

    $\alpha$-enhanced isochrones vary the temperature and luminosity of the stars at the TO point and therefore affect the hydrogen lines. According to \citet{vazdekis2001}, incorporating $\alpha$-enhanced isochrones into SPS models brings spectroscopic age estimates closer to those derived from CMDs. In order to examine the impact of $\alpha$-enhancement on the \hbo\ and \pab\ indices, we compared grids of E--MILES models with two different $\alpha$-element abundances: [$\alpha$/Fe]=0 and +0.4. The grids of these two models are shown in Fig.~\ref{fig:fig2} with dashed and dash-dotted lines, respectively. For the \hbo\ index, the grid with [$\alpha$/Fe]=+0.4 predicts only slightly higher line strengths compared to the [$\alpha$/Fe]=0 grid. In contrast, the difference is more significant in the case of the \pab\ index. The grid with [$\alpha$/Fe]=+0.4 predicts markedly higher line strengths for the \pab\ index compared to the [$\alpha$/Fe]=0 grid. This suggests that \pab\ is more sensitive to variations in $\alpha$-element abundances. Despite these differences, both $\alpha$-enhanced and solar-scaled models continue to exhibit the age zero-point problem in the optical and NIR, and unlike in \citet{vazdekis2001} for the \hg\ line, they do not help to reduce the gap. Therefore, the impact of $\alpha$-enhancement requires further investigation.

    We also include data points for MW GCs (red squares)  in Fig.~\ref{fig:fig2}  to compare the behaviour of MW GCs with that of extragalactic GCs in the context of the age zero-point problem. Similar to Cen~A GCs, most of the MW GCs fall below the E--MILES model grids for both the \hbo\ and \pab\ indices, suggesting ages that are consistently older than the oldest age of the models and even exceeding the age of the Universe itself.

    To further investigate the age zero-point problem in stellar population models, we explored the relationship between the \hbo\ and \pab\ indices. Figure~\ref{fig:fig3} illustrates this relationship for GCs from the Cen~A and MW galaxies. In the plot, Cen~A GCs are represented by filled circles, contrasting with the filled squares that represent MW GCs. The [MgFe]$^{\prime}$ metallicity values are indicated through colour bars, each assigned a specific colour for the GCs of each galaxy. We overplot the E--MILES model grid with the BaSTI isochrones. We see that the models exhibit a correlation between \hbo\ and \pab, with more variation in \hbo\ than in \pab. The greater variability in \hbo\ is expected, given the different nature of the transitions: \hbo\ originates from the Balmer series involving transitions to the n=2 level of hydrogen, while \pab\ is part of the Paschen series, involving transitions to the n=3 level. This results in distinct behaviours with age; the \pab\ strength decreases sharply at approximately 1.5~Gyr, while \hbo\ shows a more gradual decline over time. We note that contrary to the model prediction, the data reveal a significant spread in \pab,\, while \hbo\ varies within a narrower range. Additionally, no evident distinction is observed between the Cen~A and MW GCs in the plot, implying comparable ages for both galactic and extragalactic GCs. Moreover, the data do not show a clear trend with metallicity. A detailed comparison of the \hbo\---\pab\ relation among three distict SPS models is presented in Appendix~\ref{sec:AppendixA}. 

    One potential avenue to alleviate the age discrepancies in GCs is the inclusion of atomic diffusion in the isochrones. \citet{vazdekis2001} demonstrated that atomic diffusion can significantly alter the TO temperatures of low-mass stars at a relatively old age, yielding spectroscopic ages more consistent with those derived from CMDs. Furthermore, the findings of \citet{vazdekis2001} suggest that the contribution of the HB to the Balmer lines in metal-rich clusters is likely negligible, redirecting the focus towards other factors that might influence the TO temperature of main sequence stars in the models. Further investigation of this area is of paramount importance, but is beyond the scope of this paper; here we aim to show the intrinsic nature of this problem across different spectral ranges.

    It is also worth mentioning the study by \citet{schiavon2002b} on the modelling of the luminosity function (LF) at the giant branch level for better reproduction of the integrated spectrum. Their analysis shows that using observed LFs instead of theoretical ones aligns spectroscopic ages with those derived from the CMD, countering the shortcomings of theoretical LFs, which often neglect AGB stars or underestimate bright giants. According to these authors, this miscalculation, which is particularly evident in the case of 47~Tuc, has previously led to exaggerated spectroscopic ages due to the neglect of factors like He diffusion and $\alpha$-enhancement. A significant open question pertains to the applicability of the red giant excess observed in 47~Tuc ---which is a high-metallicity GGC--- across GCs of different metallicities and in extragalactic GCs with compositions distinct from those in the MW.

    Finally, it is worth considering the impact of incorporating binary stars into SPS modelling. Mass transfer and stellar interactions in binary systems within a stellar population can modify the apparent age of the population by rejuvenating stars or shortening their lifetimes. Models such as BPASS \citep{stanway2018}, which account for binary evolution, suggest that binaries may lead to younger apparent ages in observed stellar populations. However, we note that BPASS models rely on theoretical spectral libraries, which appear to have difficulty in modelling NIR spectra, particularly for cool stars and molecular bands that dominate the NIR light. The models used in our study are based on empirical spectral libraries, providing reliable predictions in the NIR.

\section{Summary and conclusions} \label{sec:conclusions}

    This paper presents a pioneering application of the extragalactic globular cluster near-infrared spectroscopy survey, wherein we extend the investigation of the model zero-point problem to the NIR range. We focus on the Pa$_{\beta}$ line strength at 1.28~microns for GCs in the Cen~A galaxy and compare this to values for the GCs in the MW. Our main findings can be summarised as follows.

    \begin{itemize}

        \item {Revealing the age zero-point problem in the NIR:} The key conclusion of our study is the recognition of the model zero-point problem in the NIR spectrum by means of the \pab\ index. Previously noted in the optical spectra of MW GCs, this discrepancy has now been identified in the NIR as the majority of the GCs in the Cen~A galaxy are observed to appear older than the predictions of the stellar population models with the oldest ages, that is, older than the current age of the Universe. Our results highlight the intrinsic nature of the age zero-point problem across different spectral ranges.

        \item {The model zero-point problem is seen in extragalactic GCs in the NIR:} The inclusion of MW GCs in this study reveals similar trends, showing that the model zero-point problem in the NIR applies to GCs outside the Galaxy.

        \item {Isochrone selection:} Our results indicate that the choice of isochrones cannot be leveraged to solve the age zero-point problem. Both BaSTI and Padova isochrones in the E--MILES model show similar inconsistencies in the optical and NIR ranges. 
        
        \item {$\alpha$-enhancement:} $\alpha$-enhanced models still exhibit the age zero-point issue in both optical and NIR ranges. Contrary to some expectations, $\alpha$-enhancement did not reduce the gap between the oldest models and GCs, highlighting the need for further investigation into its influence on stellar populations predictions.

    \end{itemize}

    In summary, our study highlights the challenge of determining GC ages using hydrogen lines. Our study presents insights into the age zero-point problem in SPS models, confirming its reach into the NIR spectrum. This problem is not limited to a specific galactic context but is a broader challenge affecting both galactic and extragalactic GCs. Our analysis shows that this issue is not resolved by varying isochrones or introducing $\alpha$-enhanced models, as both approaches still exhibit the same discrepancies. With the advent of observing facilities such as the Extremely Large Telescope and the James Webb Space Telescope, which are equipped with enhanced NIR capabilities, we have an unprecedented opportunity to observe extragalactic GCs with greater precision. Their precise data will allow refined calibrations of SPS models, improving our understanding of stellar evolution and the age zero-point problem.

\begin{acknowledgements}
      We thank the anonymous referee for their insightful comments and suggestions, which have greatly improved the quality and clarity of this manuscript. EE and AV acknowledge support from grants PID2021-123313NA-I00 and PID2022-140869NB-I00 from the Spanish Ministry of Science, Innovation and Universities MCIU. This work has also been supported through the IAC project TRACES, which is partially supported through the state budget and the regional budget of the Consejer{\'{i}}a de Econom{\'{i}}a, Industria, Comercio y Conocimiento of the Canary Islands Autonomous Community. EE and AV also acknowledge support from grant POKEBOWL PID2021-123313NA-I00 from the MCIN/AEI and the European Regional Development Fund (ERDF). RR acknowledges support from the Fundaci\'on Jes\'us Serra and the Instituto de Astrof{\'{i}}sica de Canarias under the Visiting Researcher Programme 2023-2025 agreed between both institutions. RR, also acknowledges support from the ACIISI, Consejer{\'{i}}a de Econom{\'{i}}a, Conocimiento y Empleo del Gobierno de Canarias and the European Regional Development Fund (ERDF) under grant with reference ProID2021010079, and the support through the RAVET project by the grant PID2019-107427GB-C32 from the Spanish Ministry of Science, Innovation and Universities MCIU. RR also thanks to Conselho Nacional de Desenvolvimento Cient\'{i}fico e Tecnol\'ogico  ( CNPq, Proj. 311223/2020-6,  304927/2017-1 and  400352/2016-8), Funda\c{c}\~ao de amparo \`{a} pesquisa do Rio Grande do Sul (FAPERGS, Proj. 16/2551-0000251-7 and 24/2551-0001282-6 ), Coordena\c{c}\~ao de Aperfei\c{c}oamento de Pessoal de N\'{i}vel Superior (CAPES, Proj. 0001). LGDH acknowledges support by National Key R\&D Program of China No.2022YFF0503402, and National Natural Science Foundation of China (NSFC) project number E345251001. ACS acknowledges funding from the Conselho Nacional de Desenvolvimento Científico e Tecnológico (CNPq), the Rio Grande do Sul Research Foundation (FAPERGS) and the Chinese Academy of Sciences (CAS) President's International Fellowship Initiative (PIFI) through grants CNPq-11153/2018-6, CNPq-314301/2021-6, FAPERGS/CAPES 19/2551-0000696-9, E085201009.
\end{acknowledgements}

%
%

\bibliographystyle{aa} 



\begin{appendix}

\section{Comparison of different SPS models}\label{sec:AppendixA}

    Discrepancies in GCs age predictions by SSP models in the optical have been observed in the past. \citet{mendel2007} focused on measuring ages, metallicities, and $\alpha$-element abundances for GGCs using a variety of stellar population models (namely \citealt{thomas2004, lee2005, vazdekis2010}). They found notable discrepancies in the ages predicted by different SPS models based on fitting simultaneously as many spectral indices as possible. Specifically, the average ages derived were 10.74 $\pm$ 1.84~Gyr for the \citet{thomas2004} model, 9.38 $\pm$ 1.82~Gyr for the \citet{lee2005} model, and 11.70 $\pm$ 3.60~Gyr for the \citet{vazdekis2010} model. These findings emphasise the need for additional calibration of SSP models, particularly when interpreting age--sensitive indices.

    In addition to our examination of the age zero-point problem within the framework of the E--MILES SSP models, we extend our investigation to encompass other recent stellar population models. Below, we summarise the construction and characteristics of these models:

    \begin{itemize}

        \item \citet{verro2022b} (hereafter XSL models):

            The XSL SSP models employ 639 star spectra from the X--shooter Spectral Library (XSL) DR3 \citep{verro2022a}, which covers a wide range of stellar parameters (effective temperature, surface gravity, and metallicity). These models offer a broad wavelength spectrum from the U to the K bands at moderately high spectral resolution (R$\sim$10000). The XSL SSP models employ two sets of isochrones: the latest PARSEC/COLIBRI (P/C) isochrones \citep{bressan2012, tang2014, chen2014, chen2015, marigo2013, rosenfield2016, pastorelli2019, pastorelli2020}, which provide an advanced treatment of stellar evolutionary stages, particularly the thermally pulsing asymptotic giant branch (TP-AGB) phase, and the Padova isochrones \citep{girardi2000}, which offer a simpler approach, lacking a third dredge-up and thus excluding the transition to carbon-rich TP-AGB stars. For this study, we selected a representative set of SSPs with P/C (Padova) isochrones, computed with Kroupa IMF, with ages between 5 and 15.8~Gyr (5.01 to 12.6~Gyr), whilst the total metallicity ranges from -1.4 to 0.0~dex (-1.68 to 0.00~dex).

        \item \citet{conroy2018} (hereafter C18): 
    
            We also incorporated the grids of base SSP models from the C18 SPS model and fitter ALF. These SSPs were constructed using standard isochrone synthesis techniques using the MIST series of isochrones \citep{choi2016, dotter2016}. The abundance scale of the MIST isochrones is set by the \citet{asplund2009} value of $Z_\odot = 0.0142$. The Spectral Polynomial Interpolator \citep[SPI;][]{villaume2017} was used to assign a stellar spectrum to each isochrone point. SPI was trained on the MILES and E--IRTF stellar libraries, using stellar parameters from \citep{prugniel2011} and \citep{sharma2016}. The training set was further supplemented by empirical spectra from the \citep{mann2015} M dwarf library. The C18 models offer options for flexible IMFs, including variable power--law indices and low--mass cutoffs. Specifically, for our research purposes, we utilised the C18 models with stellar ages from 5 to 13.5~Gyr and within the metallicity range of -1.5 to 0.0~dex. In order to represent the MW IMF, we adopted an IMF parametrisation with a dual--slope power--law structure: x1=1.3 for the mass range 0.08-0.5~\ms\ and x2=2.3 from 0.5 to 1~\ms. Beyond 1~\ms, the slope adheres to the Salpeter IMF \citep{salpeter1955}.

    \end{itemize}

    It is noteworthy to mention that the isochrones of all these models are solar--scaled, such that [Z/H]=[Fe/H] for the base SSPs used in this work.

\begin{figure}
        \centering
        \includegraphics[width=0.99\linewidth]{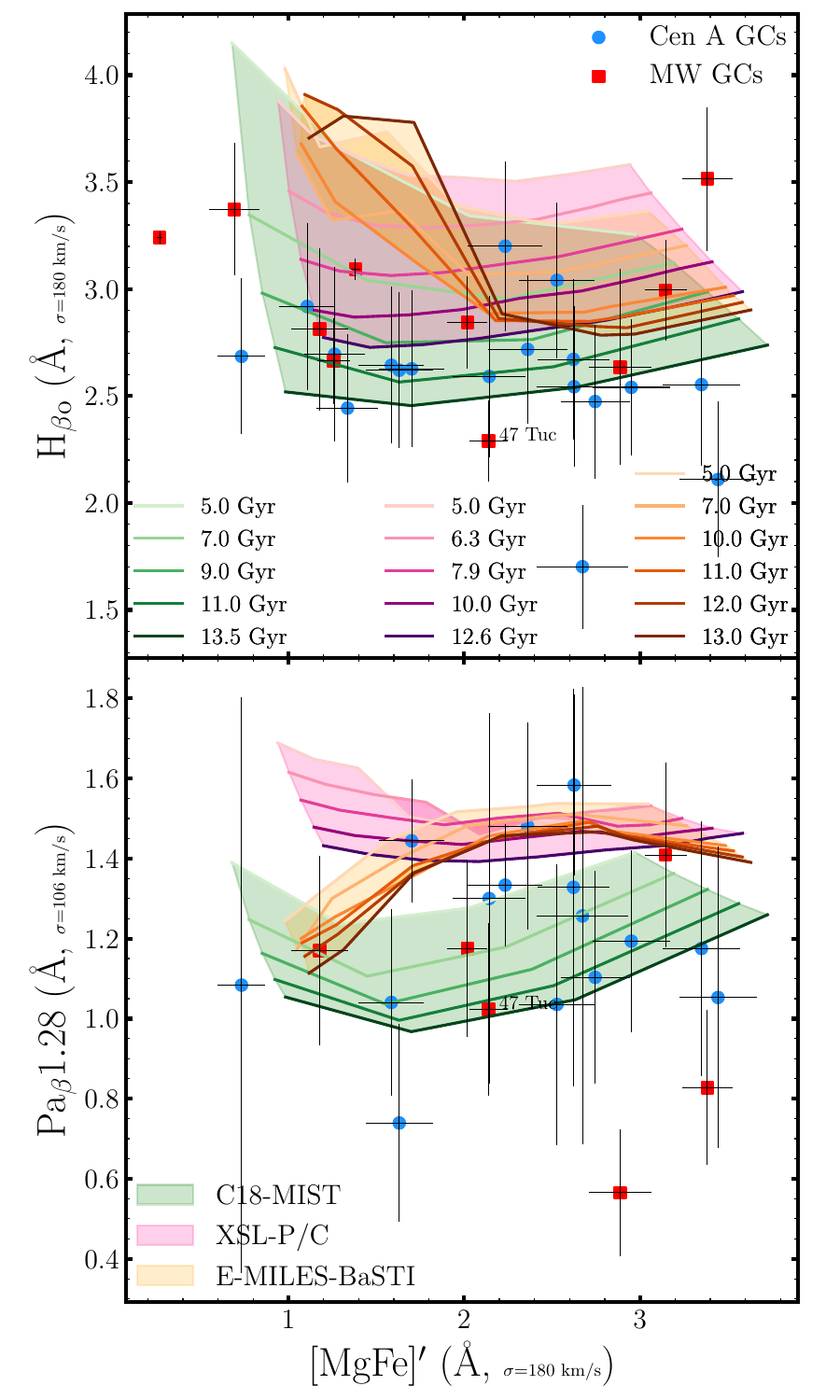}
        \protect\caption[]{\hbo\ and \pab\ line strengths of GCs from the Cen~A galaxy (blue points) and the MW (red squares) against the metallicity indicator index, [MgFe]$^{\prime}$. Predictions from various SPS models are superimposed: C18 with MIST isochrones in green, XSL with P/C isochrones in pink, and E--MILES with BaSTI isochrones in orange. Different lines within each colour represent predictions for varying ages, as the legends indicate.}
        \label{fig:figA1}
    \end{figure}

    Figure~\ref{fig:figA1} shows the \hbo\ and \pab\ line-strengths for GCs in the Cen~A galaxy (blue points), as a function of the [MgFe]$^{\prime}$ total metallicity indicator. Three distinct model grids are overlaid: E--MILES SSP model with BaSTI isochrone (in orange), XSL SSP model with P/C isochrone (in pink) and C18 with MIST isochrone (in green). The three models have different predictions for the \hbo\ index in the metal-poor regime, in part due to the differing treatment of the HB phase and mass loss. Most of the Cen~A GCs fall below the XSL model grid, similar to the E-MILES grid. However, the oldest SSPs from the C18 model predict the lowest values for \hbo, and, as a result, most of the Cen~A GCs fall within the C18 grid. Another interesting aspect of Fig.~\ref{fig:fig3} lies in the divergent predictions of the E--MILES, XSL, and C18 models in the NIR spectrum, as evidenced by the distinct shapes of their respective grids. The C18 model predicts the lowest line strengths for the \pab\ index compared to other models and encompasses nearly half of the Cen A GC sample. However, a few GCs fall above the C18 models with an age of 5~Gyr, which contrasts with the much older ages derived for these GCs by \citet{beasley2008} using optical data. Furthermore, an internal inconsistency is observed between the ages determined with these models using the \hbo\ and \pab\ panels, as the \pab\ panel generally indicates younger ages for the GCs. Similar to the E-MILES models, only a few GCs fall within the XSL grid in the \pab\ panel. Therefore, the predictions of the E--MILES and XSL models show consistency between the optical and NIR spectral ranges. Another interesting point is that, in the metal--poor regime ([MgFe]$^{\prime}$ $<$ 2.2~\AA), the E-MILES model shows contrasting behavior for the \hbo\ and \pab\ indices. As metallicity decreases, the strength of \hbo\ increases, while the strength of \pab\ decreases. This could be attributed to the different dependencies of \hbo\ and \pab\ on HB stars. In old, metal--poor ($\rm [Z/H] < -1$ dex) populations, the morphology of the HB includes stars along the blue, intermediate, and red HB. The stars contributing mostly to \hbo\ in the optical range are the blue and intermediate HB stars, which can be hotter than the TO stars, thus strengthening the \hbo\ index. Conversely, the stars contributing mostly to \pab\ in the J band are the redder HB stars, which are cooler than the TO stars, thereby weakening the \pab\ index. 

    Similar to Cen~A GCs, most of the MW GCs fall below the E--MILES model grid for both the \hbo\ and \pab\ indices, suggesting ages that are consistently older than the oldest age of the models and even exceeding the age of the Universe itself. However, the trend differs for the XSL and C18 models. For the \hbo\ index, only three MW GCs exhibit weaker absorption than the oldest SSPs in the XSL model (one in the case of the C18 model). For the \pab\ index, all reliable MW GC measurements fall below the XSL model predictions, similar to the E–MILES model. In contrast, for the C18 model, only two MW GCs fall below the model grid.

\begin{figure}
        \centering
        \includegraphics[width=0.99\linewidth]{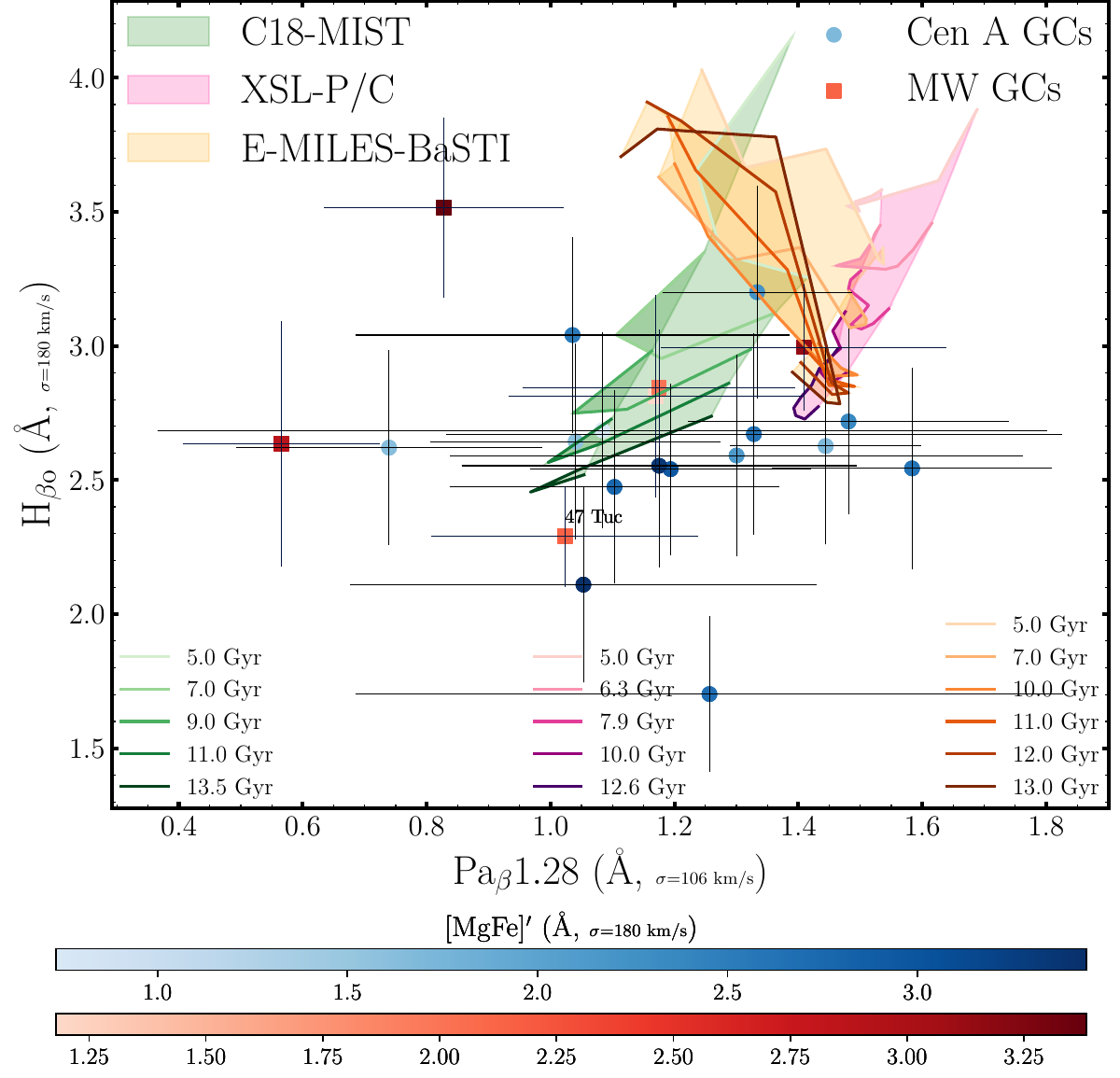}
        \protect\caption[]{Comparative analysis of \hbo\ and \pab\ line strengths in GCs from the Cen~A galaxy and the MW. In this plot, Cen~A GCs are represented by filled circles and MW GCs by filled squares. The line strengths are plotted against the [MgFe]$^{\prime}$ metallicity indicator, with a bluish colourbar denoting the [MgFe]$^{\prime}$ values for Cen~A GCs and a reddish colourbar for MW GCs. Predictive grids from three distinct models are overlaid: C18 with MIST isochrones (green), XSL with P/C isochrones (pink), and E--MILES with BaSTI isochrones (orange). Each model grid displays predictions across a range of metallicity for distinct ages specified in the legend.}
        \label{fig:figA2}
    \end{figure}

    In Fig.~\ref{fig:figA2}, we explored the relationship between the \hbo\ and \pab\ indices in different SPS models. The metallicity range of each model is as following: BaSTI-based E--MILES models, -1.49$<$Z$<$0.06, P/C-based XSL model, -1.4$<$Z$<$0.0, and MIST-based C18 model, -1.5$<$Z$<$0.0. The E--MILES model shows an increase in the \hbo\ index for the oldest SSPs with decreasing metallicity, contrasting with the behaviour of the XSL and C18 models. This discrepancy may stem from different approaches to modelling HB morphology in metal--poor SSPs. In older metal--poor populations, the prevalence of hot HB stars leads to increased Balmer line strengths, causing these populations to appear younger. The complex interplay of factors like mass loss, metallicity, and dynamical effects makes detailed HB morphology modelling challenging. The XSL and C18 models likely have employed a more conservative mass loss approach, resulting in cooler evolved stars, unlike the E--MILES model.

\section{Comparison of models with different isochrones}\label{sec:AppendixB}

\begin{figure}
        \centering
        \includegraphics[width=0.99\linewidth]{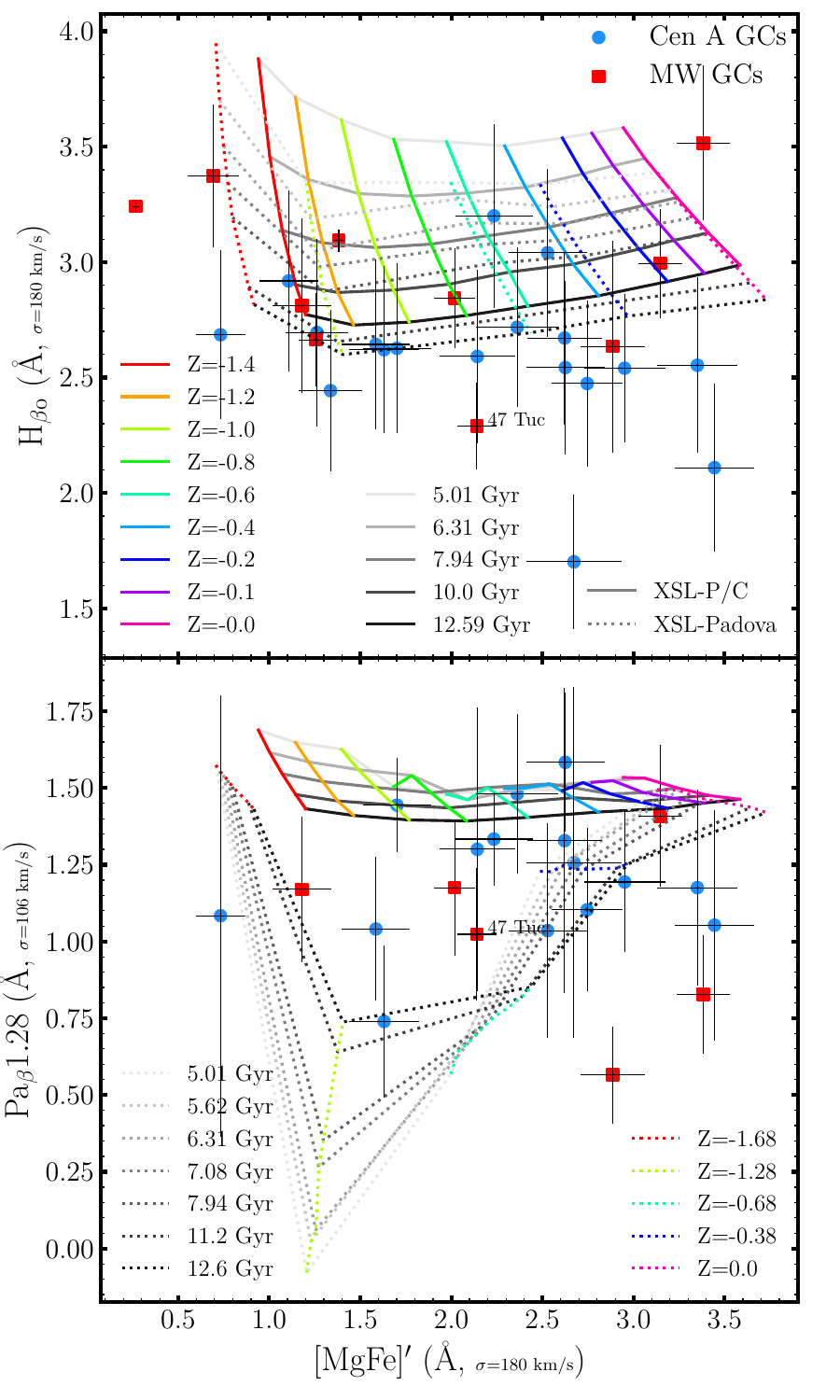}
        \protect\caption[]{Comparison of different isochrones. \hbo\ (upper panels) and \pab\ (lower panels) line strengths of GCs from the Cen~A galaxy (blue points) and the MW (red squares) are plotted against the metallicity indicator [MgFe]$^{\prime}$. XSL models with P/C and Padova isochrones are shown with solid and dotted grids, respectively.}
        \label{fig:figB}
    \end{figure}

    In Fig.~\ref{fig:figB}, we compared two different sets of isochrones for the XSL model: PARSEC/COLIBRI (grid with solid lines) and Padova (grid with dotted lines). It is important to mention that the C18 model is based solely on one set of isochrones and is therefore not included in this analysis. The upper panel of Fig.~\ref{fig:figB} shows model predictions for the \hbo\ index versus the [MgFe]$^{\prime}$ indicator, and the lower panel shows the same but for the \pab\ index. For both indices, the grid based on the Padova isochrones consistently lies lower than the one based on the P/C isochrones. The oldest SSP with the Padova isochrone predicts a sudden increase in the \hbo\ line-strength at Z $<$ -0.6, whereas the oldest SSP with the P/C isochrone shows this increase at a lower metallicity of -1.0 dex. Similar to P/C isochrone, the XSL model with Padova isochrone shows the age zero-point problem too.
    
    A particularly notable difference arises in the shape of the \pab\ grid when using the Padova-based XSL model. In the lower panel of Fig.~\ref{fig:figB}, the Padova grid exhibits a completely different shape compared to the P/C grid, and most of the GCs in our sample fall within the Padova grid. This result contrasts with the behavior of the E--MILES model, where the Padova isochrone does not show significant differences from the BaSTI grid for the \pab\ index (see Fig.\ref{fig:fig2}). The consistency between the E--MILES grids across different isochrones provides more stable results. 

\end{appendix}

\end{document}